# Entanglement-assisted electron microscopy based on a flux qubit

Hiroshi Okamoto[a)] and Yukinori Nagatani[b)]

[a)]*Department of Electronics and Information Systems, Akita Prefectural University, Yurihonjo 015-0055, Japan*
[b)]*National Institute for Physiological Sciences, Okazaki 444-8787, Japan*

A notorious problem in high-resolution biological electron microscopy is radiation damage to the specimen caused by probe electrons. Hence, acquisition of data with minimal number of electrons is of critical importance. Quantum approaches may represent the only way to improve the resolution in this context, but all proposed schemes to date demand delicate control of the electron beam in highly unconventional electron optics. Here we propose a scheme that involves a flux qubit based on a radio-frequency superconducting quantum interference device (rf-SQUID), inserted in essentially a conventional transmission electron microscope. The scheme significantly improves the prospect of realizing a quantum-enhanced electron microscope for radiation-sensitive specimens.

[a)] Electronic mail: okamoto@akita-pu.ac.jp

It is important to use a minimal number of electrons when imaging a vitrified biological specimen in cryoelectron microscopy [1, 2], in order not to destroy high-resolution features [3]. However, this entails a large amount of shot noise due to the particle nature of the electron. Recently, several quantum schemes emerged to address this problem [4-9]. These schemes involve, e.g., a ring-shaped electron optics [7] or a Cooper pair box placed on a cryogenic electron mirror [9]. In this Letter, while based on the previous concept of entanglement-assisted transmission electron microscopy (TEM) presented in terms of a charge qubit [9], we propose to significantly accelerate the development of it by the use of a magnetic alternative, namely an rf-SQUID based flux qubit.

These quantum approaches are distinct from low-voltage TEM (LVTEM) [10, 11] that also aims at reducing radiation damage, which, however, may not be suitable for imaging relatively thick (~ 30 nm) hydrated biological specimens containing large macromolecules of biological interest. In addition, the theoretical performance of our approach exceeds that of LVTEM and in-focus phase contrast TEM [12].

The proposed TEM scheme is a confluence of several major ideas of the past. First, the earliest discussion [13] on macroscopic quantum coherence (MQC) was in terms of the rf-SQUID, where both the clockwise and counter-clockwise supercurrents can exist in quantum superposition. Second, the Aharonov-Bohm effect [14, 15] has been convincingly demonstrated [16] in a TEM, using magnetic flux confined in a loop of a superconducting tube. Finally, recently emerged entanglement-assisted quantum metrology helps us to beat the standard quantum limit [17].

First, we briefly review how entanglement-assisted TEM works. Biological specimens consist of light atoms and behave as weak phase objects. Suppose that we want to detect the difference $\Delta\varphi$ of phase shifts associated with two small neighboring specimen regions $S_0$ and $S_1$.

Repetition of such measurements will eventually produce an image. Let the probe electron states $|0\rangle$ and $|1\rangle$ be ones that respectively pass the specimen regions $S_0$ and $S_1$. Define the symmetric/antisymmetric states $|s\rangle = (|0\rangle + |1\rangle)/2^{1/2}$ and $|a\rangle = (|0\rangle - |1\rangle)/2^{1/2}$, respectively. The probe electron wave passing $S_1$ acquires the phase difference $\Delta\varphi$ relative to the one going through $S_0$. In the conventional case, the measurement boils down to distinguishing states $(|0\rangle + |1\rangle)/2^{1/2} = |s\rangle$ and $(e^{-i\Delta\varphi/2}|0\rangle + e^{i\Delta\varphi/2}|1\rangle)/2^{1/2} = \cos(\Delta\varphi/2)|s\rangle - i\sin(\Delta\varphi/2)|a\rangle \sim |s\rangle - (i\Delta\varphi/2)|a\rangle$. The former state correspond to the specimen that does not induce phase shift difference between the two states $|0\rangle$ and $|1\rangle$, while the latter does. If a measurement basis $\{|s\rangle, |a\rangle\}$ is used, clearly it takes $N \sim (2/\Delta\varphi)^2$ electrons to detect the phase difference. In contrast, we use a qubit incorporated in the illumination part of the TEM, with available states $|0\rangle_q$ and $|1\rangle_q$, that interacts with the probe electron in entanglement-assisted TEM. As shown below, we obtain a state $(e^{-ik\Delta\varphi/2}|0\rangle_q + e^{ik\Delta\varphi/2}|1\rangle_q)/2^{1/2}$ on the qubit after using a group of $k$ electrons. In this case, it takes $(2/k\Delta\varphi)^2$ such groups, or $N' \sim k(2/k\Delta\varphi)^2 = (1/k)(2/\Delta\varphi)^2$ electrons to detect the phase difference. In particular, if we are able to let $N' = k$, then we attain the Heisenberg limit $N' \sim 2/\Delta\varphi$. (In actual imaging, we use a different measurement basis [9] because the sign of $\Delta\varphi$ needs to be determined. However, the simplified argument here suffices to illustrate the quantum advantage.)

The following procedure produces the above-mentioned state $(e^{-ik\Delta\varphi/2}|0\rangle_q + e^{ik\Delta\varphi/2}|1\rangle_q)/2^{1/2}$ with $k$ electrons. Initially, the qubit is set to the symmetric state

$$|s\rangle_q = (e^{-i\sigma/2}|0\rangle_q + e^{i\sigma/2}|1\rangle_q)/2^{1/2}, \quad (1)$$

where $\sigma = 0$. The initial electron wave produced by an electron gun interacts with the qubit to generate an entangled state $(e^{-i\sigma/2}|0\rangle|0\rangle_q + e^{i\sigma/2}|1\rangle|1\rangle_q)/2^{1/2}$. The specimen induces a phase shift $\Delta\varphi$ to the electron state $|1\rangle$ relative to $|0\rangle$ upon transmission and the state of the system becomes

$$(e^{-i(\sigma + \Delta\varphi)/2}|0\rangle|0\rangle_q + e^{i(\sigma + \Delta\varphi)/2}|1\rangle|1\rangle_q)/2^{1/2}. \quad (2)$$

Finally, the electron is detected in the far field (or any plane conjugate to a plane where the diffraction pattern is observed. Let us call such plane a *diffraction plane*.) by an area detector. Suppose that the electron is detected on the $j$-th pixel of the area detector. Let the state of the electron be $|d_j\rangle$ and write $\langle d_j|0\rangle = a_j$ and $\langle d_j|1\rangle = b_j$. These coefficients are known to the experimenter. We expect not to be able to distinguish the electron states $|0\rangle$ and $|1\rangle$ by such a measurement because the detector is placed in the far field. This means that $|a_j| = |b_j|$ holds to a good approximation because the two states $|0\rangle$ and $|1\rangle$ generates a similar intensity pattern on the area detector, which in turn allows us to write $a_j = b_j e^{-i\beta_j}$, where $\beta_j$ is a known phase angle. Therefore, by 'capping' Eq. (2) with $\langle d_j|$ we obtain, up to the overall phase factor

$$(e^{-i(\sigma + \beta_j + \Delta\varphi)/2}|0\rangle_q + e^{i(\sigma + \beta_j + \Delta\varphi)/2}|1\rangle_q)/2^{1/2}. \quad (3)$$

Clearly, the above process starting with the state (1) is valid also with non-zero initial $\sigma$. Hence, by repeating the process $k$ times without resetting $\sigma$ we obtain

$$(e^{-i(\sigma + \Sigma\beta_j + k\Delta\varphi)/2}|0\rangle_q + e^{i(\sigma + \Sigma\beta_j + k\Delta\varphi)/2}|1\rangle_q)/2^{1/2}, \quad (4)$$

where $\Sigma\beta_j$ symbolically represents sum of all the phase angles $\beta_j$ (with generally different pixel number $j$'s) obtained in the course of the measurement. Since $\Sigma\beta_j$ is known, one can manipulate the qubit to cancel this phase to obtain the desired state $(e^{-ik\Delta\varphi/2}|0\rangle_q + e^{ik\Delta\varphi/2}|1\rangle_q)/2^{1/2}$ and proceed to measuring it.

In the previously proposed scheme [9], a charge qubit based on a Cooper pair box [18] is used. This scheme, while sound in principle, has several major complications when it comes to practical aspects. Figure 1 (a) shows the scheme. The electron emitted from the electron gun first goes through a monochromator and then directed to a cryogenic electron mirror by a beam separator. The probe electron is slowed down by an electron mirror, on which the charge qubit is placed, to electro-statically interact with the qubit. After interacting with the qubit, the electron goes through the beam separator for the second time and is accelerated up to several hundred keV towards the specimen. The first difficulty is the necessity for exquisite electron beam monochromaticity. The electron mirror repels the electron by an electric field, which, if too strong, would prevent us from 'seeing' a small electrostatic potential change on the surface. In order to detect the minute potential change of the order of ~ 400 μV that reflects the charge state of the Cooper pair box, the electric field must be kept at ~ 3 kV/m [9]. This is in contrast to the usual value of $10^7$ V/m. This implies that the electrons with slightly lower kinetic energy cannot even approach the Cooper pair box. It has been shown that the energy spread of the electron beam must be of the order of 1 meV [9]. This kind of value has been achieved only in the context of high-resolution electron energy loss spectroscopy [19], where electron kinetic energy is on the order of 1 eV. Second, the above considerations strongly suggest that the portion of electron optics containing the electron gun, monochromator, beam separator, and the electron mirror (encircled by a dotted line in Fig. 1 (a)) should best be kept at a similar electrostatic potential to maintain a small kinetic energy of electrons throughout in the portion. However, this means that the whole portion of the electron optics that includes the Cooper pair box, with associated dilution refrigerator, low-noise control electronics and microwave lines, must be floated at near the electrostatic potential of the electron gun, which is several hundreds of kilovolts (and is negative). While this is not a fundamental problem, it does represent a significant practical problem. Third, the scheme, as presented in Ref. [9], employs a Cooper-pair box qubit with an energy difference between the low-energy symmetric state and the high-energy antisymmetric state of ~ 1 GHz in terms of frequency. This means that the state vector on the Bloch sphere representing the qubit state rotates with that frequency. Consequently, electron emissions from the electron gun must be synchronized with this motion, necessitating the use of a pulsed electron source, presumably driven by a pulsed laser synchronized with qubit control electronics.

We intend to address the above problems by the use of a magnetic alternative to the Cooper pair box, which is an rf-SQUID flux qubit. First of all, we propose to place the SQUID qubit in the high-energy electron beam (See Fig 1 (b)) so that there is no need to electrically float a large portion of the electron microscope, and furthermore eliminates the necessity for an unusually good monochromator. It is clear that magnetic elements handle high-energy electrons more effectively than electrostatic elements because the Lorenz force is proportional to the velocity of the electron. Consider a situation, where the waist of a Gaussian electron beam with the waist size $a$ passes through a single magnetic flux quantum $\phi_0 = h/2e$, which is trapped in a superconducting rectangular ring with a dimension $a$ by $l$, where the side with the length $l$ is along the optical axis. Upon passing, the electron beam experiences a Lorenz force $F = evB = ev\phi_0/al$. The time duration for this is $\Delta t \sim l/v$ and we have the beam deflection angle $\theta_d = \Delta p/p \sim h/2pa$. On the other hand, the diffraction-limited beam has the angular beam spread $\theta_b \sim \lambda/a$, where $\lambda = h/p$ is the electron wavelength. To produce a distinct quantum state of the electron beam, the deflection angle should not be 'buried' in the beam spread due to diffraction. Here we have a constant $\theta_d/\theta_b$ ratio, of the order 1, independent of the electron velocity. (In contrast, when a charge qubit is used, the deflection angle is estimated to be $\theta_d = \Delta p/p \sim e^2/\varepsilon_0 avp$, which is far too small for our purpose.) Nevertheless, the magnetic field distribution near a SQUID ring is not simple, entailing difficulties in computing electron-optical characteristics associated with it. Moreover, if the above ratio $\theta_d/\theta_b$ turns out to be considerably smaller than 1 in such computations, we are in trouble.

A device based on the Aharonov-Bohm (AB) effect offers a 'clean' solution to the above problem. In this case, the magnetic flux quantum is confined in a superconducting hollow ring, so that the trapped magnetic flux forms a closed line. The hollow ring is placed in a plane normal to the optical axis. Then, the electron wave propagating through inside the loop acquires a phase shift $\pi$ when compared to the wave passing outside the ring, if a single magnetic flux is trapped inside the ring. Let us denote this state by $|1>_q$, anticipating this to be a qubit state. (No phase shift occurs when no magnetic flux is trapped. We write this state $|0>_q$.) Let the electron state that corresponds to the wave going outside the ring be $|s>$, and the one corresponding to the wave propagating through the ring be $|a>$, respectively. Let these states be properly normalized. Then, an initial electron state $|0> = (|s> + |a>)/2^{1/2}$ will transform to itself, or the state $|1> = (|s> - |a>)/2^{1/2}$, depending on the existence of magnetic flux quantum inside the ring. These two states are orthogonal and are suitable for the purpose of entanglement-assisted TEM.

For the above AB-effect device to work, the magnetic flux confined inside the hollow ring must be in quantum superposition. Such a macroscopic superposition can be realized by using an properly flux-biased rf-SQUID circuitry, the concept originally introduced in the context of MQC [13]. Since our rf-SQUID must have the unusual hollow-ring geometry, it may seem to place an excessive demand on our micro-fabrication capability. This, however, is not necessarily the case. In the past several decades, the fabrication technology in the field of micro-electromechanical systems (MEMS) has advanced to enable etching of fifty-to hundreds of μm deep trenches with 90 +- 1 degree sidewalls in silicon [20]. Figure 2 shows a possible geometry of the device. The scale we have in mind is, e.g. a thousand μm thick silicon wafer, on which features are microfabricated with a lateral size on the order of 10 μm. However, this size depends on the spatial coherence of the electron beam we can get in the TEM. A central hole and a ring-shaped hole (interrupted with a silicon beam) are drilled with deep reactive ion etching (DRIE). The silicon beam mechanically supports the ring-shaped silicon mass, in which the magnetic flux is trapped. Superconducting material is deposited on all vertical walls and the bottom surface (not shown) of the ring-shaped silicon mass, forming a system akin to a short-circuited coaxial cable. Two Josephson junctions are lithographically fabricated on the upper surface of the wafer, through which the magnetic flux quantum moves in or out. A weak, separately applied magnetic field modulates the effective Josephson energy of the combined parallel Josephson junctions in the standard manner.

For an rf-SQUID to manifest the MQC effect, a condition $Li_c \sim \phi_0$ is usually required, where $L$ is the inductance of the rf-SQUID and $i_c$ is the critical current of the (combined) Josephson junction. A rough estimate of $L$ may be obtained if we view the ring-shaped silicon mass as a 'coaxial cable'. We have, neglecting a logarithmic factor, $L \sim \mu d$ where $\mu$ is magnetic permeability and $d$ is the thickness of the silicon wafer. Approximately 1 μA critical current is appropriate, assuming $d = 1$ mm. In general, the lateral size of an rf-SQUID qubit is quite large (~ 100 μm) to obtain a sufficiently large inductance [21]. In our case, the lateral size should be made smaller than the coherent electron wave front. Hence we propose to extend the size along the optical axis.

In an actual rf-SQUID qubit, the difference of the magnetic flux $\Delta\phi$ between the two quantum states $|0>_q$ and $|1>_q$ is typically smaller than $\phi_0$. Although we believe that $\Delta\phi$ can be made close to $\phi_0$, we point out that if it is difficult, one could in principle use a multiple-turn, as opposed to a single-turn, magnetic flux loop providing that such microfabrication is possible.

In principle, adjustability of the effective Josephson energy allows for ample freedom for operation [22]. For example, one might tune the barrier height between the two minima in the rf-SQUID potential landscape to realize a sufficiently slow MQC oscillation frequency of, e.g. ~ MHz, between the states $|0>_q$ and $|1>_q$. In this case, the qubit state is essentially still compared to the dynamics of probe electrons during the measurement. Hence, instead of using a pulsed electron source with precisely synchronized timing, a much simpler beam blanker (similar to ones employed in e-beam lithography systems) could be used to let approximately $k$ electrons pass the specimen at once (e.g. in ~ 10 nS) to enable entanglement-assisted TEM. Clearly, however, a detailed analysis of qubit operation needs to be carried out in a future study.

We briefly describe electron optics, starting from the upstream of the electron beam. First, in order not to let the electron beam impinge on the solid part of the rf-SQUID qubit, a stencil mask (an example pattern is shown in Fig. 3 (a)) is provided in a diffraction plane. At a following plane conjugate to the specimen plane (call them image planes), an aperture is provided to remove the high spatial frequency part of the electron beam on a diffraction plane. Hence, the electron beam intensity on the following diffraction plane is 'smoothed', as shown in Fig. 3 (b). The qubit is located on this plane. Since the qubit should be cooled down to millikelvins, a (possibly side-entry) dilution refrigerator is employed. To avoid heating due to infrared radiation, multiple shielding is required and the solid angle to high temperature parts seen from the qubit must be minimized with low-temperature apertures. The electron source should have good spatial coherence so that the wave front is coherent over the entire qubit with the size ~ 10 μm. Then, the condenser lens focuses the resultant electron beam onto the specimen. Figures 3 (c) and 3 (d) shows the electron beam intensity on the specimen when the qubit is in the state $|0>_q$ and $|1>_q$, respectively. Hence we detect the phase-shift difference between these two illuminated parts on the specimen. These parts should be within the 'delocalization length' of inelastic scattering to avoid measurement errors [9]. Finally, in the far field we detect the scattered electron with a high-efficiency area detector [23]. There, we see the 'shadow' of the SQUID qubit. Depending on whether the electron is detected inside the shadow of the SQUID ring or not, the phase compensation angle $\beta_j$ is either 0 or $\pi$.

This research is supported in part by the JSPS Kakenhi grant #25390083.

FIGURES AND CAPTIONS

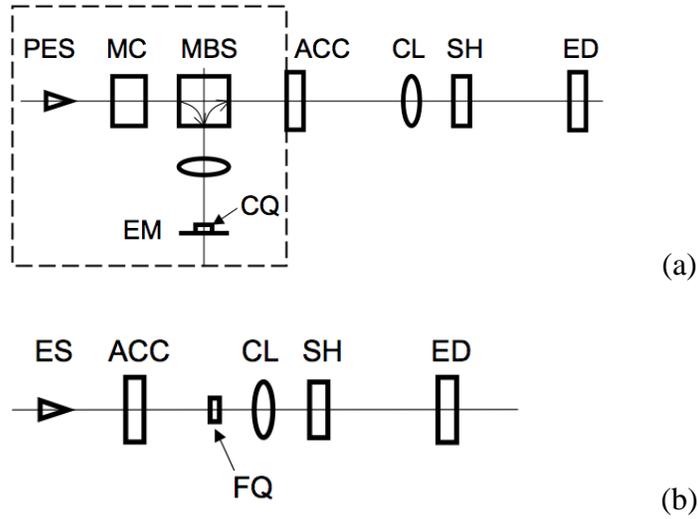

(a)

(b)

Figure 1: Structures of entanglement-assisted TEMs. (a) A scheme based on a charge qubit (CQ), which is placed on an electron mirror (EM) connected to a magnetic beam separator (MBS). The portion encircled by the dotted line is floated to a large electrostatic potential. (b) A scheme based on a flux qubit (FQ) that is simply inserted to a TEM. These schemes include an electron source (ES) or pulsed electron source (PES), monochromator (MC), accelerator (ACC), condenser lens (CL), specimen holder (SH), and electron area detector (ED). Not all the lenses are shown.

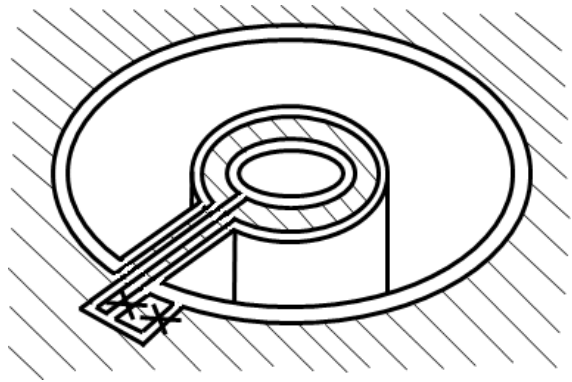

Figure 2: A possible structure of the flux qubit device. Hatched parts are insulating (e.g., silicon dioxide), while white parts are made of superconductor. Two "X" symbols represent Josephson junctions.

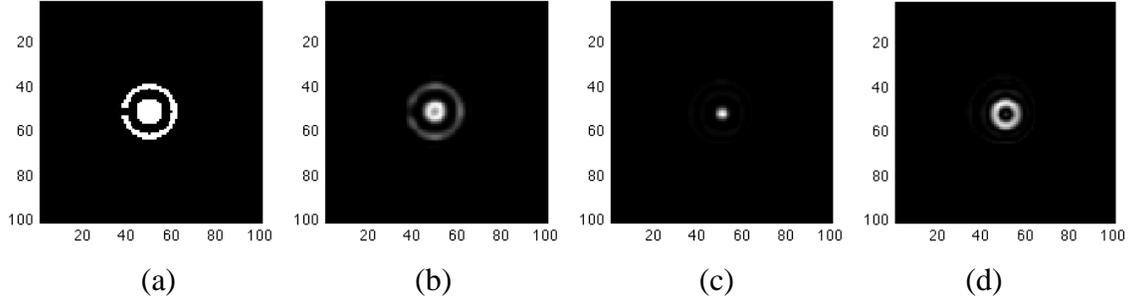

(a) (b) (c) (d)

Figure 3: Numerical simulation of the electron optics. (a) The stencil mask pattern. (b) The electron beam shape incident on the SQUID qubit. (c) An electron beam intensity map on the specimen when the qubit is in the state $|0>_q$. (d) A corresponding map when the qubit state is $|1>_q$. The approximate sizes of these patterns are supposed to be ~ 10 μm for (a) and (b), and a few nm for (c) and (d). However, these patterns are simply Fourier transformed to each other in the computation and the actual sizes can be set rather arbitrarily, using different excitations of electron lenses. In this simulation, we assume that the magnetic flux associated with the states $|0>_q$ and $|1>_q$ are well defined and the difference between them is $\phi_0$.